\documentclass[a4paper,11pt]{article}

\usepackage[authoryear, sort]{natbib}
\setcitestyle{authoryear,round,semicolon,aysep={,}}
\usepackage[T1]{fontenc}
\usepackage[english]{babel}
\usepackage{graphicx}
\usepackage{xcolor}
\usepackage{booktabs}
\usepackage{tabularx}
\usepackage[nointegrals]{wasysym}
\newcommand{\fullc}{\CIRCLE}
\newcommand{\halfc}{\LEFTcircle}
\newcommand{\emptyc}{\Circle}
\newcommand{\nac}{\textup{--}}
\usepackage{authblk}
\usepackage[margin=2.5cm]{geometry}
\usepackage{caption}
\usepackage[colorlinks=true,urlcolor=black,linkcolor=black,citecolor=black]{hyperref}

\usepackage{titlesec}
\titlespacing*{\section}
  {0pt}{2.5ex plus 1ex minus .2ex}{1.5ex plus .2ex}
\titlespacing*{\subsection}
  {0pt}{2ex plus 1ex minus .2ex}{1ex plus .2ex}

\title{\vspace{-2cm}\textbf{Reclaiming the Social in Social Media}}
 
\author[1]{Luca Benn}
\author[1,2]{Moritz Merz}
\author[1,3,4,5]{David Grüning}

\affil[1]{Scent (in formation), Germany.}
\affil[2]{Harvard University, Cambridge, USA.}
\affil[3]{University of Cambridge, Cambridge, UK.}
\affil[4]{Max Planck Institute for Human Development,\protect\linebreak Center for Adaptive Rationality, Berlin, Germany.}
\affil[5]{Stanford University, Stanford, USA.}

\date{}

\begin{document}

\maketitle

\vspace*{-1cm}
\begin{center}
\small \textbf{Author contacts:}  \href{mailto:luca.benn@scent.social}{luca.benn@scent.social};
\href{mailto:moritz.merz@scent.social}{moritz.merz@scent.social}/\href{mailto:mmerz@g.harvard.edu}{mmerz@g.harvard.edu}; \href{mailto:david.gruning@mrc-cbu.cam.ac.uk}{david.gruning@mrc-cbu.cam.ac.uk}/\href{mailto:dgruening@mpib-berlin.mpg.de}{dgruening@mpib-berlin.mpg.de}/\href{mailto:gruening@stanford.edu}{gruening@stanford.edu}
 
\smallskip
\end{center}
\vspace*{-0.4cm}
\noindent\rule{\linewidth}{0.4pt}

\begin{abstract}
\noindent
Concerns about polarization, antisocial behavior, and mental health have broadly led to two responses to social media: adapting platforms through moderation and prosocial design, and restricting access to these platforms through age limits and comparable measures. Both address consequences of social media while leaving the attention-driven architecture intact. We argue that the harms commonly attributed to social media arise not from technologies supporting social connection but from their implementation within the attention economy. In response, we ask what a digital social environment built for genuine human connection would look like. From four principles -- \textsc{Purpose}, \textsc{Alignment}, \textsc{Transparency}, and \textsc{Access} -- we can derive four technical properties that fulfill these principles: \textit{Operator Blindness}, \textit{Algorithmic Sovereignty}, \textit{Operational Parity}, and \textit{Verifiability}. To the best of our knowledge, no deployed alternative satisfies all four. We propose an architecture that does, making the attention economy not just discouraged but structurally impossible. We explain how the resulting user experience refocuses on each user's individual social environment and differs from that of current platforms.

\smallskip
{\noindent
\par}
\end{abstract}

\section{Two Responses, One Blind Spot}\label{sec1}
Social media have become central to modern society. Their role is especially pronounced among younger generations, for whom they are nearly ubiquitous spaces for communication, entertainment, identity exploration, and civic engagement \citep{auxier2021, boulianne2020young, vogels2022}. This centrality has also raised concern among researchers, policymakers, and practitioners about a multitude of consequences, such as polarization, antisocial behavior, and declining mental health \citep{cheng2015antisocial, kubin2021role, orben2019, surgeongeneral2023social}.

Responses to these concerns have largely taken two forms. One adapts the problematic environments, seeking to align platforms with human flourishing by adding features such as content moderation or prosocial design 
\citep{gruening2025} or by rearranging the engagement system (e.g., alternative platforms like Mastodon, Bluesky, or Sparkable). The other restricts access altogether, particularly for vulnerable groups, as reflected in recent age-limit legislation in Australia and the UK, and similar debates in other countries. However much these two approaches have their unique strengths, we argue that they share one blind spot. Both largely address the consequences attributed to existing social media architectures, while the attention-driven design that produces them remains intact.

In this paper, we aim to contribute a more fundamental rethinking of what social media can be. Rather than a new design within existing technical paradigms, we propose a new paradigm in which the attention economy is structurally precluded, so that social media are social by necessity. We are currently implementing an architecture that realizes this paradigm as a prototype. A full protocol specification and technical discussion are to be released separately \citep{benn_inprep}.

\section{Engagement Over Connection}\label{assessment}

We argue that many of the negative effects attributed to social media share a common root. Their origin lies not in the concept of ``social media'', but in its implementation through the attention economy. Today, this distinction has become difficult to draw, as, over time, major platforms have largely converged on the same attention-driven architecture.

Social media platforms originally served the social affordances of the internet, yet over time shifted largely toward maximizing time spent by users. Most recently, this shift has intensified through AI being deployed as an engagement generator and a content creator. Users now connect less with people they know offline than with distant content creators, thereby forming parasocial relationships characterized by one-sided intimacy \citep{horton1956}. Many scholars accordingly question whether these platforms can still be called ``social'' media 
\citep{boyd2026,toernberg2026}. 

How far the harms attributed to social media extend remains the subject of ongoing debate \citep{valkenburg2022,orben2019,odgers2020}. Our argument does not depend on resolving this debate. It requires only that the attention-driven architecture undermines the social purpose of these platforms, as illustrated by the shift toward parasocial relationships.

What we face is a fundamental problem, one that demands fundamental rethinking. This paper is an invitation to reconsider, from the ground up, what digital technology designed for human social flourishing could look like. 
Can an architecture restore the purpose of social media by making its exploitation for attention technically \textbf{impossible}?

\section{Rethinking the Fundamentals}\label{sec:rethinking}

\begin{quote}
    \centering \itshape If we attached no associations to the term ``social media'', \\ how would we design a digital social environment for genuine human connection?
\end{quote}
We propose the following principles as the foundation for a reconceived social media technology. They articulate the values underlying this technology and constitute a set of requirements that, to the best of our knowledge, no existing solution meets to date.

\subsection{Principles of a Humane Technology}\label{sub:principles}

A social media technology exists to fulfill the \textsc{\bfseries Purpose} for which it is named: the formation of directed, genuine connections between people. Platforms have not yet sustained this \textsc{Purpose} against competing, often economic, incentives.

\smallskip

The system therefore requires \textsc{\bfseries Alignment}: its incentives and \textsc{Purpose} must coincide. This precludes earning from attention or data, since this demonstrably tends to produce structural interests that diverge from those of users.

\smallskip

Nonetheless, contrary incentives can develop within complex systems, and blind trust in the operator would leave them undetected. These conditions demand \textsc{\bfseries Transparency}. The system's architecture, code, and cryptographic foundations must be open to inspection, so that its properties are independently verifiable.

\smallskip

To fulfill \textsc{Purpose}, the solution must grant \textsc{\bfseries Access} to everyone.
Technical expertise or dedicated infrastructure, for instance, can exclude many users. Participation must demand nothing but being human.

\subsection{Where They Lead}\label{sec:properties}

\begin{figure}[t]
    \centering
    \includegraphics[]{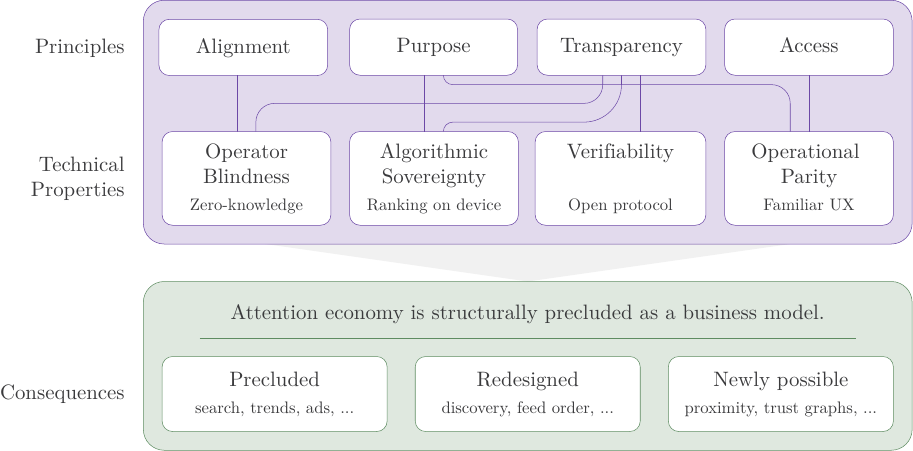}
    \caption{Our proposed principles are fulfilled by a set of technical properties that specify how they can be implemented in practice. With this technology, the attention economy is impossible, and many platform-defining features are either precluded or redesigned, while entirely new ones become feasible.}
    \label{fig:fig1}
\end{figure}

Once we accept these principles, the question is which technical properties would be sufficient to meet them (see Figure~\ref{fig:fig1}).

\smallskip

\textsc{Alignment} permits no incentive that could undermine the system's \textsc{Purpose}, and \textsc{Transparency} requires this to be independently verifiable. A promise not to misuse data satisfies neither condition, since data that exists will always attract competing interests, and promises rest on blind trust. \textit{\bfseries Operator Blindness} is accordingly the only verifiable guarantee: the platform operator is structurally unable to know who the platform's users are, who knows whom, or what passes between them.

\smallskip

Algorithmically personalized feeds have defined social platforms' architecture for over a decade. \textit{Operator Blindness} precludes them entirely, since the data required to compute personalized relevance is unavailable to the operator. Fulfilling \textsc{Purpose} nonetheless demands an experience users find compelling, with content selected and ordered around individual relevance. \textit{\bfseries Algorithmic Sovereignty} resolves this tension through \textsc{Transparency}, which requires ranking to be determined by the user, on their own device, under transparent rules they control.

\smallskip

A compelling user experience also requires a low entry barrier. The requirements of \textsc{Access} preclude achieving \textit{Operator Blindness} through a fully decentralized system, since operating personal infrastructure is a barrier for many users. This can be solved with a hybrid architecture: a blind server stores encrypted containers under meaningless labels, ensuring availability, reliability, and speed, while access to content is managed exclusively on users' own devices. \mbox{\textit{\bfseries Operational Parity}} ensures that, for the user, the solution is no more than a smartphone app, with straightforward profile creation and onboarding, key custody, and recovery from key loss despite \textit{Operator Blindness}. Relevant content is available from the first session and social discovery is supported. Latency remains low, and effective action against abusive content and users is structurally possible.

\smallskip

\textsc{Transparency} demands that trust in the system's behavior rest on \textit{\bfseries Verifiability}. The protocol must accordingly be public, the implementation open-source, and every cryptographic primitive standardized and widely deployed.

\medskip 

Together, these four technical properties are sufficient to meet our principles. In Table~\ref{tab:comparison}, we have assessed established open-source platforms that meet \textit{Verifiability} based on their compliance with the remaining properties. None of them combines all three. 

Since we have derived these properties as a sufficient set rather than a necessary one, a missing property does not imply that a platform does not meet our principles. However, as the table illustrates, the missing property in each case is directly related to difficulties the system faces in scaling or in fulfilling the principles. \textit{Operator Blindness}, together with \textit{Operational Parity}, is met only by platforms without a feed, among which Signal combines a high level of privacy with the user experience one would expect from a messaging app. Therefore, our solution satisfying all properties is intended to be a viable and \textsc{Purpose}-driven alternative to contemporary social media.

\begin{table}[t]
  \centering
  \small
  \setlength{\tabcolsep}{5pt}
  \renewcommand{\arraystretch}{1.25}
  \begin{tabularx}{\linewidth}{@{}p{0.12\linewidth} cccc >{\raggedright\arraybackslash}X@{}}
    \toprule
    \textbf{System} & \textbf{OB} & \textbf{AS} & \textbf{OP} & \textbf{VE}
      & \textbf{Principles at risk} \\
    \midrule
    Mastodon / ActivityPub & \emptyc & \halfc  & \halfc  & \fullc &
      \textsc{Alignment}: operator can read content and graph. \newline\textsc{Access}: avoiding operators requires self-hosting. \\
    Bluesky / AT Protocol  & \emptyc & \halfc  & \fullc  & \fullc &
      \textsc{Alignment}: public repositories, concentrated hosting. \\
    Nostr                  & \emptyc  & \fullc  & \halfc & \fullc &
      \textsc{Alignment}: public notes let aggregators rank engagement. \newline\textsc{Access}: raw key custody without recovery. \\
    Peergos & \fullc & \halfc & \emptyc & \fullc &
      \textsc{Access}: no iOS app. \newline\textsc{Purpose}: minimal feed within file storage, no social discovery. \\
\midrule
    \multicolumn{6}{@{}l}{Messengers without a feed (\textsc{Purpose} narrowed):}\\
    Matrix & \halfc & \nac & \fullc & \fullc &
      \textsc{Alignment}: homeserver can see membership, metadata. \\
    Signal & \fullc & \nac & \fullc & \fullc &
      Blindness achieved only without a feed. \\
    Briar & \fullc & \nac & \emptyc & \fullc &
      \textsc{Access}: peers must be online, no recovery. \\
    \bottomrule
  \end{tabularx}
  
    \smallskip

    \renewcommand{\arraystretch}{1.2}
    \begin{tabularx}{\linewidth}{@{}l l
    >{\hsize=1.2\hsize\centering\arraybackslash}X
    >{\hsize=0.9\hsize\centering\arraybackslash}X
    >{\hsize=0.9\hsize\centering\arraybackslash}X@{}}
    \toprule
    & \textbf{Property} & \fullc~\textbf{satisfied} & \halfc~\textbf{partial} & \emptyc~\textbf{not met} \\
    \midrule
    OB & Operator Blindness      & content \& graph hidden & graph exposed    & content in plaintext \\
    AS & Algorithmic Sovereignty & open \& transparent        & limited or opaque     & operator-imposed \\
    OP & Operational Parity      & app, recovery \& discovery  & one major hurdle & several hurdles \\
    VE & Verifiability           & open \& standardized    & partly open      & closed \\
    \bottomrule
  \end{tabularx}
  \caption{Deployed open-source systems against the four technical properties (top), with the rubric for each rating (bottom).
    No system attains all three of \textit{Operator Blindness}, \textit{Algorithmic Sovereignty}, and \textit{Operational Parity}. The ratings are based on the properties derived here, not on the objectives of the respective projects, and reflect designs as of September 1, 2026.}
  \label{tab:comparison}
\end{table}

\medskip

\textit{Operator Blindness} and \textit{Algorithmic Sovereignty} form the foundation of human-centered social media. They also structurally exclude the attention economy as a viable business model, an intended consequence that nonetheless raises the question of financial sustainability. Any for-profit model faces the challenge of remaining in \textsc{Alignment} with the system's declared \textsc{Purpose}. Donation-based funding, as practiced by Signal and Wikipedia, and public or institutional funding, as in the case of Mastodon, demonstrate that such a platform can be viable as a non-profit enterprise. Financial sustainability and structural integrity are therefore not in conflict.

\newpage
\section{Functional Implications}

Because the proposed system's architecture diverges so fundamentally from that of existing platforms, the social media experience changes with it. The following examples illustrate what this means in practice.

\medskip
\textbf{Some familiar functions are structurally precluded.} 
Since the operator is blind, features that require centralized content categorization cannot be offered: there is no content search bar, no interest-based feed training, and no ``suggested accounts'' or ``people you might like'' feature based on behavioral profiles. Trending topics, viral charts, or system-wide rankings of popular content cannot be displayed either. Consequently, the platform is not an open public square for mass discourse or clashing views. Instead, content and connections reach a user exclusively through their own social environment. The feed is finite and can end. Strangers cannot message, comment, or otherwise appear in a user's space unless they are introduced by an existing connection, and content spreads no further than its author allows. \textit{Operator Blindness}, \textit{Algorithmic Sovereignty}, and \textit{Verifiability} together preclude tracking-based advertising. Users encounter no sponsored posts, promoted content, or commercial placements.

\medskip
\textbf{Other functions can be redesigned.} 
To enable the social discovery required by \textit{Operational Parity} despite \textit{Operator Blindness}, content must reach the feed via the user’s own social network. Which interactions lead to content distribution, what types of content exist, and how much social metadata is transmitted with them together shape the evolving social dynamics.
\textit{Algorithmic Sovereignty} lets the resulting feed be sorted chronologically, by locally computed relevance, or by other modes based on user preference. What should happen when the feed ends? Removing the attention economy makes it possible to ask whether the visibility of likes and follower counts is needed at all. The platform type is shaped by whether connections stay strictly mutual or admit one-way follows, and whether content carries a limited lifespan. Addressing abuse requires that users disclose reported content to the operator. How far such disclosure reaches, and what evidence the operator then relies on, remain open to discussion.

\medskip
\textbf{Entirely new functions become possible for the first time.}
Sensitive data can travel fully end-to-end encrypted and be evaluated only on the user's own device. Features that extractive architectures previously foreclosed now become available. This includes, for example, location-aware and proximity-based features, trust propagation within social circles, and novel cryptographic protocols for group coordination.

\section{Looking Forward}

The harms commonly attributed to social media arise not from technologies supporting social connection but from the attention-economy architecture on which today’s platforms have converged. From four principles of a fundamentally social platform -- \textsc{Purpose}, \textsc{Alignment}, \textsc{Transparency}, and \textsc{Access} -- we derived four technical properties that are jointly sufficient to satisfy them -- \textit{Operator Blindness}, \textit{Algorithmic Sovereignty}, \textit{Operational Parity}, and \textit{Verifiability} -- and showed how they transform the user experience.
Together, these properties constitute an architecture in which the attention economy is structurally precluded, so that its features serve each user's own social environment.

To demonstrate feasibility, we are implementing the architecture as a prototype and will publish the protocol specification along with a complete technical description \citep{benn_inprep}.

What we offer here is a starting point for fundamentally rethinking social media technology rather than a finished product. We explicitly invite readers to contribute to this rethinking and to sharpen a shared vision of what social media could become.

\section*{Acknowledgments}

David Grüning’s work is funded by the Huo Family Foundation and Stanford’s Center for Digital Health.
The \textit{Scent} project is registered at the Student Project House (SPH) of ETH Zürich, which provided funding, coaching, and workspace. The e-na'bel Foundation contributed technology funding. Aki Zürich provided additional workspace.

\section*{Competing Interests}

The authors are developing the system described in this paper. The funders acknowledged above had no role in the design of this work, the writing of this paper, or the decision to publish.

\section*{Contributions}

Conceptualization: LB, MM, DG. Structural logic of the system: LB, MM. Technical implementation: LB, MM. Supervision: DG. Writing the first draft: LB, MM, DG. Revising further drafts: LB, MM, DG.

\newpage
\bibliographystyle{plainnat}
\bibliography{references}

@techreport{auxier2021,
  author      = {Auxier, Brooke and Anderson, Monica},
  title       = {Social Media Use in 2021},
  institution = {Pew Research Center},
  address     = {Washington, DC},
  year        = {2021},
  month       = apr,
  note        = {Retrieved August 8, 2026, from https://www.pewresearch.org/internet/2021/04/07/social-media-use-in-2021/}
}

@techreport{vogels2022,
  author      = {Vogels, Emily A. and Gelles-Watnick, Risa and Massarat, Navid},
  title       = {Teens, Social Media and Technology 2022},
  institution = {Pew Research Center},
  address     = {Washington, DC},
  year        = {2022},
  month       = aug,
  note        = {Retrieved August 8, 2026, from https://www.pewresearch.org/internet/2022/08/10/teens-social-media-and-technology-2022/}
}

@article{boulianne2020young,
  author  = {Boulianne, Shelley and Theocharis, Yannis},
  title   = {Young people, digital media, and engagement: A meta-analysis of research},
  journal = {Social Science Computer Review},
  volume  = {38},
  number  = {2},
  pages   = {111--127},
  year    = {2020},
  doi     = {10.1177/0894439318814190}
}

@inproceedings{cheng2015antisocial,
  author    = {Cheng, Justin and Danescu-Niculescu-Mizil, Cristian and Leskovec, Jure},
  title     = {Antisocial behavior in online discussion communities},
  booktitle = {Proceedings of the International {AAAI} Conference on Web and Social Media},
  volume    = {9},
  pages     = {61--70},
  year      = {2015},
  doi       = {10.1609/icwsm.v9i1.14583}
}

@article{kubin2021role,
  author  = {Kubin, Emily and von Sikorski, Christian},
  title   = {The role of (social) media in political polarization: A systematic review},
  journal = {Annals of the International Communication Association},
  volume  = {45},
  number  = {3},
  pages   = {188--206},
  year    = {2021},
  doi     = {10.1080/23808985.2021.1976070}
}

@article{orben2019,
  author  = {Orben, Amy and Przybylski, Andrew K.},
  title   = {The association between adolescent well-being and digital technology use},
  journal = {Nature Human Behaviour},
  volume  = {3},
  number  = {2},
  pages   = {173--182},
  year    = {2019},
  doi     = {10.1038/s41562-018-0506-1}
}

@techreport{surgeongeneral2023social,
  author      = {{Office of the Surgeon General}},
  title       = {Social Media and Youth Mental Health: The {U.S.} {Surgeon General}'s Advisory},
  institution = {U.S. Department of Health and Human Services},
  address     = {Washington, DC},
  year        = {2023},
  url         = {https://www.hhs.gov/sites/default/files/sg-youth-mental-health-social-media-advisory.pdf}
}

@misc{gruening2025,
  author = {Gr{\"u}ning, David J. and Kamin, Julia},
  title  = {Prosocial Design in Trust and Safety},
  year   = {2025},
  doi    = {10.48550/arXiv.2506.12792},
  note   = {arXiv:2506.12792. Book chapter, forthcoming in: Trust and Safety: Past, Present, and Future}
}

@unpublished{benn_inprep,
  author = {Benn, Luca and Merz, Moritz and Gr{\"u}ning, David J.},
  title  = {Scent: An Operator-Blind Architecture for Social Media},
  note   = {Manuscript},
  year   = {in preparation}
}

@article{horton1956,
  author  = {Horton, Donald and Wohl, R. Richard},
  title   = {Mass communication and para-social interaction: Observations on intimacy at a distance},
  journal = {Psychiatry},
  volume  = {19},
  number  = {3},
  pages   = {215--229},
  year    = {1956},
  doi     = {10.1080/00332747.1956.11023049}
}

@article{boyd2026,
  author  = {{boyd}, {danah}},
  title   = {Social Media Is Now Parasocial Media},
  journal = {Social Media + Society},
  volume  = {12},
  number  = {2},
  pages   = {20563051261437487},
  year    = {2026},
  doi     = {10.1177/20563051261437487}
}

@misc{toernberg2026,
  author = {T{\"o}rnberg, Petter and Rogers, Richard},
  title  = {Towards a Post-Social Media Studies},
  year   = {2026},
  note   = {SocArXiv preprint, not peer reviewed. Retrieved August 8, 2026, from https://osf.io/preprints/socarxiv/6nue7}
}

@article{valkenburg2022,
  author  = {Valkenburg, Patti M. and Meier, Adrian and Beyens, Ine},
  title   = {Social media use and its impact on adolescent mental health: An umbrella review of the evidence},
  journal = {Current Opinion in Psychology},
  volume  = {44},
  pages   = {58--68},
  year    = {2022},
  doi     = {10.1016/j.copsyc.2021.08.017}
}

@article{odgers2020,
  author  = {Odgers, Candice L. and Jensen, Michaeline R.},
  title   = {Annual Research Review: Adolescent mental health in the digital age: facts, fears, and future directions},
  journal = {Journal of Child Psychology and Psychiatry},
  volume  = {61},
  number  = {3},
  pages   = {336--348},
  year    = {2020},
  doi     = {10.1111/jcpp.13190}
}

\end{document}